\documentclass[twocolumn,preprintnumbers,amsmath,amssymb]{revtex4}
\usepackage{graphicx}
\usepackage{dcolumn}
\usepackage{bm}

\newcommand{\be}{\begin{equation}}
\newcommand{\bel}[1]{\begin{equation}\label{#1}}
\newcommand{\ee}{\end{equation}}
\newcommand{\bea}{\begin{eqnarray}}
\newcommand{\ba}{\begin{array}}
\newcommand{\eea}{\end{eqnarray}}
\newcommand{\ea}{\end{array}}

\begin{document}

\title{\bf Computational modelling of the collective stochastic motion of Kinesin nano motors  }

\author{Yousef Jamali $^{1}$ , M. Ebrahim Foulaadvand $^{1,2}$ and H. Rafii Tabar $^{1}$ }

\affiliation{ $^1$ Computational Physical Sciences Research Laboratory,
Department of Nano-Science, Institute for Research in Fundamental Sciences (IPM), P.O. Box 19395-5531, Tehran, Iran.\\
$^2$ Department of Physics, Zanjan University, P.O. Box 45196-313, Zanjan, Iran.}

\begin{abstract}

We have developed a two dimensional stochastic
molecular dynamics model for the description of intra
cellular collective motion of bio motors, in
particular Kinesins, on a microtubular track. The model is capable
or reproducing the hand-over-hand mechanism of the directed motion
along the microtubule. The model gives the average directed velocity and the
current of Kinesins along the microtubule. It is shown
that beyond a certain density of Kinesins, the average
velocity and current undergo notable decrease which is
due to formation of traffic jams in the system.

\end{abstract}

\maketitle
\section{{Introduction}}

Various complex functions of Eucariotic cells namely mitosis,
intracelluler transports and cell motility are based on the
cooperative motion of molecular motor proteins such as Dyneins,
Kinesins and Myosins \cite{howard,schliwa1,schliwa2,ajdari}.
These molecular motors move on cytoskeletal filaments like
microtubles and F-actins. The mentioned phenomena exhibit a
variety of self-organized processes and patterns which are
characterized by their time scales
\cite{schad1,nedelec1,nedelec2,schnitzer,okada}. A large number
of in-vitro experiments have revealed that the motion of Kinesin
is {\it processive} i.e., it performs stroke-type steps on its
track before detachment \cite{1p,2p,1}. It is now well
established that the Kinesins motion is derived by the free
energy released in the chemical hydrolysis reaction of ATP
(adenosine-triphosphate). The fuel of the motion that is the ATP
molecules arrive randomly in the vicinity of Kinesins.
Consequently, the directed motion takes place on a stochastic
grounds. During each step in which one period of the microtubule,
hereafter referred to as MT, is covered, one ATP molecule is
hydrolysed. The underlying mechanism responsible for this forward
motion is partially understood. Two basic mechanisms have been
proposed. In the first one the so-called {\it inchworm
mechanism}, one head of Kinesin drag the other one which is
always at the rear \cite{2}. In this mechanism, there is an
asymmetry between the lagging. The second mechanism which is
called {\it hand-over-hand} has much resemblance to human
walking. This mechanism involves repeated docking and undocking
of both heads upon breaking of the chemical bond with the MT in
the course of ATP hydrolysis \cite{3,4}. At each stepping event,
the rear head executes a forward motion of 16 nm and attaches to
the tubulin binding site ahead of the front head. Recent
empirical evidences overwhelmingly have established that
hand-over-hand mechanism is the true mechanism \cite{3}. In
recent years various approaches have been proposed to model the
directed motion of nano-sized bio motors. Simple random walk was
used as a simple though efficient framework for modelling such
kind of motion \cite{klump1,klump2,klump3,klump4}. In this
framework, Kinesins are assumed to be point particles which
execute a directed random walk on a discrete one dimensional
filament, as a model for microtubular track, with lattice spacing
equal to Kinesin walking distance i.e. $8$ nm. Due to
stochasticity in the Kinesin motion, it would be reasonable to
incorporate degrees of randomness in the modelling approach. In
this spirit, a new approach namely asymmetric particle hopping
models in continuous time which are based on master equation
approach came into play \cite{klump2,klump5,kolomeisky1,frey}.
Another theoretical approach for the modelling of bio motors
motion is the {\it so-called} ratchet mechanism \cite{ajdari}. In
this mechanism, the directed motion is generated by a
time-dependent periodic potential which is switched between two
states \cite{ajdari94,viscek,kolomeisky2,klump01,10,6,9,jamali}.
The stochasticity comes in the manner in which the potential
switches from one state to the other one and reflects the random
arrival of ATP molecules and conformational transitions which
changes the interacting potential between Kinesin and MT. In
principle, more than one Kinesins can simultaneously move on a
single MT. Kinesins do certainly interact with each other. While
the true interaction potential between adjacent Kinesins has not
yet empirically explored, in-vitro observations have revealed the
collective motion of Kinesins which arises from the interaction
between these motors \cite{schad2}. In order to find some insight
into the nature of the Kinesins motion, we have developed a
Langevin-type model which incorporates interaction among
Kinesins. Our aim is to improve our understanding on the
spatio-temporal organisations of intra cellular transport
phenomena and find the mechanism responsible for formation of
molecular traffic jams.

\section{Modelling of Kinesins movement}

The model we have considered is a generalisation of the model proposed in
\cite{jamali} for the motion of a single Kinesin.

We shall now explain our stochastic ratchet model in some details. The force acting on each Kinesin head consists of four terms as follows:

\be
{\bf F}=-\vec{\nabla} H_{ratchet} -\vec{\nabla} H_{bistable} - \vec {\nabla} H_{rep}(\Delta r) + {\bf \Gamma}.
\ee

$ H_{ratchet}$ is the stochastic ratchet potential which models the interaction between Kinesin and microtubule. $H_{bistable}$
is a bistable potential that models the elastic coupling between two heads of Kinesin, $H_{rep}$ is the repulsive potential between adjacent
Kinesins and finally ${\bf \Gamma}$ represents both the stochastic Brownian and the frictional forces experienced by Kinesin heads due
to thermal noise in the intra cellular environment. Let us explain these four terms in more details.

\subsection{ Ratchet potential}

The ratchet potential $H(x,y,t)$ should involve an asymmetry along the MT. More precisely, we have taken the form of ratchet potential as follows
\cite{jamali}:

\be
H(x,y,t)= CH_x(x)H_y(y)A(t)S(t)
\ee

With periodicity in $x$ direction i.e., $H(x+L,y,t)=H(x,y,t)$ where $L$ is the microtubule period which is about $8~nm$.
For $H_x(x)$ we have used the first eleven terms of a Fourier series of an asymmetric function $W(x)$ as follows:

\be
H_x(x) =a_0 + \sum_{m=1}^5 a_m cos(\frac{2\pi mx}{L}) + b_m sin(\frac{2\pi mx}{L})
\ee

The coefficients $a_0, a_m$ and $b_m ~~ m=1,\cdots,5$ are Fourier coefficients. For the
asymmetric function $W(x)$ we chose the saw-tooth function:\\

$$~~W(x)=\frac{10}{9}(\frac{x}{L}-[\frac{x}{L}]) ~~~~ \frac{x}{L}-[\frac{x}{L}]<0.9$$
\be
W(x)=10(\frac{x}{L}-[\frac{x}{L}])-10 ~~~ \frac{x}{L}-[\frac{x}{L}]>0.9
\ee

Where $[~]$ denotes the integer part. The summation form of $H_x(x)$ has been used for computational convenience.
Figure (1) shows $H_x(x)$ versus $x$ .

\begin{figure}
\centering
\includegraphics[width=7.5cm]{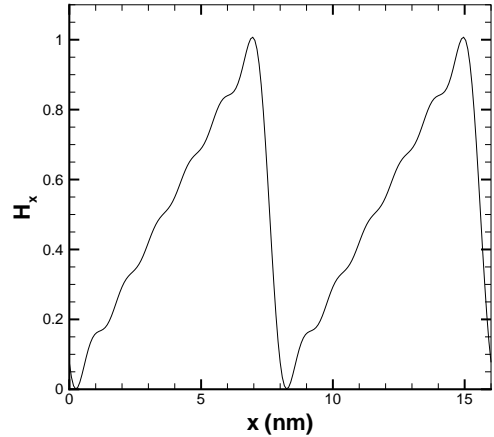}
\caption{ Periodic dependence of $H_x$ on $x$. } \label{fig:bz2}
\end{figure}

Each peak of $H_x(x)$ is located at a distance $0.1L$ from the
position of the next minimum. For more details see Ref.
\cite{jamali}. Concerning the part of potential $H_y(y)$
responsible
for the perpendicular direction to the MT we have used the following potential :\\

\be
H_y(y)= -\frac{1}{y^2+y_0^2}
\ee

where $y_0=0.1~\sqrt L$. It is assumed that the above form of
$H_y(y)$ can model the interaction between the
molecules making up the Kinesin head and the MT. The symmetry of
the function allows for detaching both in upward and downward
directions perpendicular to the MT. This function is smoother than
the Morse potential therefore we can choose larger time steps
which can speed up the simulation. Figure (2) shows $H_y$.

\begin{figure}
\centering
\includegraphics[width=7.5cm]{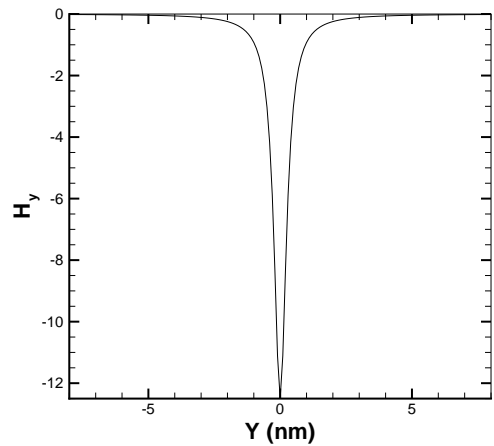}
\caption{ Dependence of $H_y$ on the perpendicular direction $y$. } \label{fig:bz2}
\end{figure}

Now we shall discuss the time dependent terms i.e., $A(t)$ and
$S(t)$ of the Ratchet potential. The stochastic variable $A(t)$
specifies the biochemical affinity of each Kinesin head. It is a
flashing potential which can switch between two discrete values $A(t)= 0,1$.
The value of $A(t)$ depends on the biochemical status of the ATPase binding
site of the head. The biochemical cycle is divided into four
status.\cite{3,12,13} In the first status, denoted by K, the
ATPase binding site contains no ATP molecule. When an ATP
molecules arrives at the binding site, the head is said to be in its
second status which is denoted by K.ATP. The third status is
associated with the hydrolysis of the absorbed ATP into ADP plus
an inorganic phosphate molecule Pi. This status is denoted by
K.ADP.Pi. Eventually after releasing the Pi molecule, the head
is characterized by its fourth status K.ADP. The
biochemical cycle is accomplished by releasing the ADP molecule.
In the status 1-3 the head has a high affinity ($A=1$) to attach,
or to keep its attached condition to the MT. On the other
hand, in the fourth status i.e., K.ADP the affinity will
considerably reduced ($A=0$) due to some conformational changes.
The second term $S(t)$ is related to the interaction between
Kinesins. Besides biochemical status, each head can be either
attached to the MT or detached from it. Each head can attach only to
particular sites on the MT. Since only one head can attach to
each MT binding site, the potential seen by each head depends on
the occupancy of its adjacent MT binding site. For adjacent Kinesins to
MT, the stochastic variable $S(t)$ describes the occupancy of the
corresponding nearest binding site on the MT. It is zero if this site is
already occupied by another head and one otherwise (see Fig.4 ). Parameter $C$
is set at $C=0.01~ eV$. The value of constant $C$ determines the
depth of the potential well. Its size must be such that the energy difference
between the on-state (A=1) and the off-state (A=0) under the equilibrium
condition is comparable with the energy released from the
hydrolysis of the ATP i.e.,

\be
[H(x,y,t)]_{A(t)=1} - [H(x,y,t)]_{A(t)=0} \leq E_{ATP \rightarrow ADP + Pi}
\ee

\subsection{Viscous and stochastic thermal noise terms}

The force ${\bf \Gamma }$, which for simplicity acts in two dimension,
has the following structure:

\be
{\bf \Gamma}_x=-\eta \dot{x}(t) + {\bf \xi}_x (t).
\ee
\be
{\bf \Gamma}_y=-\eta \dot{y}(t) + {\bf \xi}_y (t).
\ee

The first term corresponds to energy dissipation due to viscosity of the surrounding fluid. $\xi_x$ and $\xi_y$ are
stationary random Gaussian white noises with zero mean. More precisely we have:

\be
\langle \xi_i(t) \xi_j(t') \rangle = 2 \eta k_B T\delta(t-t')\delta_{ij}
\ee

The drag coefficient $\eta$ for a Kinesin in an aqueous environment is reported
at $\eta \approx 6 \times 10^{-8} pNs/nm$ \cite{viscek}. We have taken this value for $\eta$
in this paper.

\subsection{ The bistable potential}

The bistable potential is responsible for the elastic coupling of the two Kinesin heads. This potential
is expressed by the following form \cite{jamali,9}:

\be
H_{bistable}(\Delta r) = C_1[1 + (\frac{\Delta r}{l})^4 -2(\frac{\Delta r}{l})^2 ]
\ee

In which $\Delta r$ is the distance between the two heads, $C_1=0.12~eV$ is the amplitude of the potential
and represents the coupling strength between heads, and $l=0.7L$ is the distance between the two potential minima.
In the following figure, we have sketched the dependence of the bistable potential versus distance.

\begin{figure}
\centering
\includegraphics[width=7.5cm]{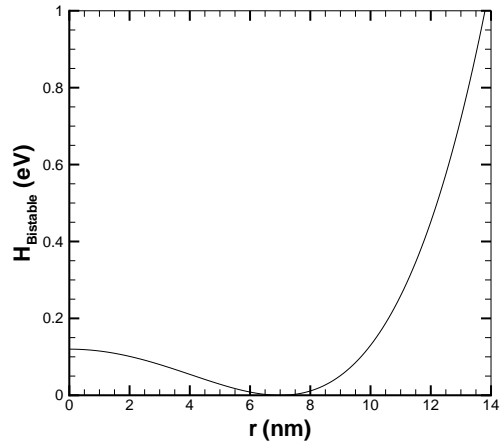}
\caption{ Bistable potential between two heads vs distance. It models the elastic interaction between Kinesin heads. } \label{fig:bz2}
\end{figure}

\subsection{ The potential between different Kinesins heads }

In order to model the interaction between adjacent Kinesins, we introduce a short-ranged repulsive potential
between heads of different Kinesins. More concisely, we have used the shifted-force repulsive part of the
Lennard-Jones potential as follows:

\be
H_{rep}(\Delta r) = U(\Delta r) - U(\Delta r_c)
\ee

In which $U(\Delta r)= \epsilon (\frac{\sigma}{r})^6$ with $\epsilon=0.2~eV,~\sigma= 70~A^{0}$ and the cut off length
$r_c=80~A^{0}$. This repulsive potential prevents the heads of adjacent Kinesins from getting too close to each other.

\subsection{simulation details}

We have not yet discussed in details how quantities $A$ an $S$ vary with time. To this end, we introduce a quantity $a(t)$ for
each Kinesin's head which specifies whether at time $t$ the head is attached to the MT or not. $a(t)$ is one if the head is attached to a
binding site and zero otherwise. Before proceeding further let us specify when we consider a head attached to a MT binding site.
If at any time step, a head has a high affinity ($A(t)=1$) and is located within a rectangular box with size $\delta_x=3.8~nm$ and
$\delta_y=0.8~nm$ centred at a binding site of the MT then we consider this head as an attached one.

\begin{figure}
\centering
\includegraphics[width=7.5cm]{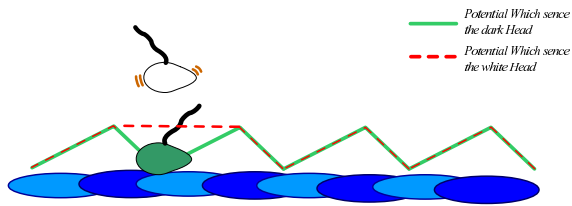}
\caption{ For adjacent heads to MT, The
ratchet potential is zero if the nearest binding site on the MT is already occupied by another head.
A head is considered attached to a binding site if it is in a small vicinity of that site. } \label{fig:bz2}
\end{figure}

Not we turn into $A(t)$. Apparently the biochemical cycles of each
head can be described by transition rates from one state to the
other one. We assume that the average time that each head spends
in the states K, K.ATP or K.ADP.Pi is in the order of $ms$ when
the head is attached to the MT \cite{14,15}. We note that this
average time depend on the attachment of the head and the
position of its partner head. The average time a head spends in
the fourth state K.ADP is much shorter and is in the order of
$\mu s$ \cite{14,16}. Since the rate of ATP consumption is
considerably reduced when a head is detached \cite{17,18,19}, in
our model we have assumed that only an attached head received an
ATP molecule. Moreover, we experimentally know the rate at which
an attached head makes a transition from the state K
$\rightarrow$ K.ATP strongly depends on the attaching state and
position of its partner head \cite{19,20}. In case the partner
head is attached and rear, the above rate is lowered about two
orders of magnitude. We have tuned the corresponding rates in our
code such that the processivity and the average velocity of a
single Kinesin coincides with the known empirical findings.
Consequently, we used these adjusted rates in the collective
motion of Kinesins. Periodic boundary condition have been imposed
and the Kinesins motion are assumed, for simplicity, to be
restricted to two dimension along $x$ and $y$ axes. The size of
the simulation box was taken as follows: $L_x=50~L,~L_y=40~L$. The
MT is located along the $x-$ axis at $y=0$ and $y \in
[\frac{-L_y}{2},\frac{L_y}{2}]$. Time step $\delta t$ was set to
$5\times 10^{-10} s$. The system has been simulated for
$3\times10^{10}$ timesteps in each run. The first $5\times10^{9}$
timesteps are discarded for reaching to equilibrium. Kinesin
global density $\rho_g$ is defined as the number of Kinesins per
binding site of the MT. Certainly $\rho_g$ can exceed one.

\section{simulation results}

Generically the Kinesin movement can be classified into 3 states:
resting, forward proccesive movement and detachent/atachement to
the MT. Once a Kinesin leaves the box from the vertical boundaries
at $y=\pm\frac{L_y}{2}$ its corresponding image re enters from
the opposite side. Those Kinesins which leave the last site in
the $x-$ direction at $i=L_x$ re enter into the MT from the left
site $i=1$. Figure (5) exhibits the trajectory of a typical
Kinesin at the global density $\rho_g=0.4$:

\begin{figure}
\centering
\includegraphics[width=7.5cm]{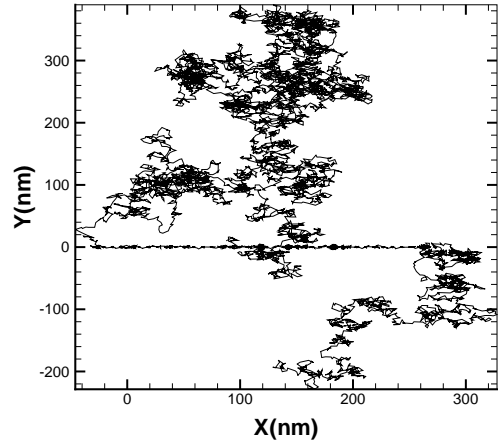}
\caption{ 2D trajectory of a typical Kinesin at $\rho_g=0.4$. Its corresponding space-time plot is sketched in Fig.7. } \label{fig:bz2}
\end{figure}

Figure (6) exhibits the space-time plots of some more Kinesins
and the time dependence of their centre of mass at $\rho_g=0.4$.
We observe that although each Kinesin undergoes rapid fluctuations
upon becoming unbound, the centre of mass (CM) has a smooth
increasing behaviour in time. We have found the CM velocity by
fitting a linear line to the the curve as $v_{cm}= 0.5 ~\mu m/s $.As can be seen, when a Kinesin is attached to the MT, its $x$ component undergoes small fluctuations and is
frequently increasing with time. It rarely undergoes backward motion which is consistent with
in-vitro experiments \cite{schnitzer,22,23}.

\begin{figure}
\centering
\includegraphics[width=7.5cm]{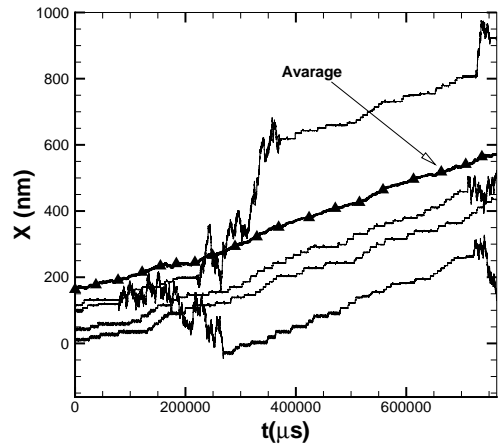}
\caption{ Space-time plots of some Kinesins
and the time dependence of their centre of mass at $\rho_g=0.4$.
Despite each Kinesin undergoes rapid fluctuations upon becoming
unbound, the centre of mass (CM) has a smooth
increasing behaviour. } \label{fig:bz2}
\end{figure}

 On the occasion of detachment from the MT, the Kinesin performs a stochastic
random motion, in the cytoplasmic environment, which is characterized by large fluctuations. It is a well-known fact that
Kinesins are processive motors i.e., they can move directedly along the filament
for relatively large distances before detaching from it. Our model can successfully reproduce this
behaviour in the collective motion of Kinesins. To illustrate this behaviour, we have explicitly shown $x$ component
time dependence of the two heads as well as the centre of mass for a typical Kinesin in the following figure.

\begin{figure}
\centering
\includegraphics[width=7.5cm]{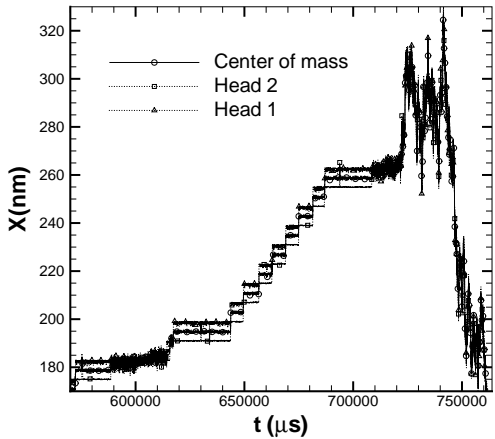}
\caption{ Processive motion of a Kinesin along the MT. } \label{fig:bz2}
\end{figure}

As can be seen from Figure (7), the hand-over-hand mechanism of
the directed motion is evident. In order to find a better
insight, we have evaluated the dependence of a single Kinesisn
mean squared displacement (MSD) on time. The MSD has been
obtained by averaging over the trajectories of all the Kinesins.
Figure (8) depicts the time behaviour of a single Kinesin' MSD
for various global Kinesin concentration $\rho_g$ in a log-log
plot.

\begin{figure}
\centering
\includegraphics[width=7.5cm]{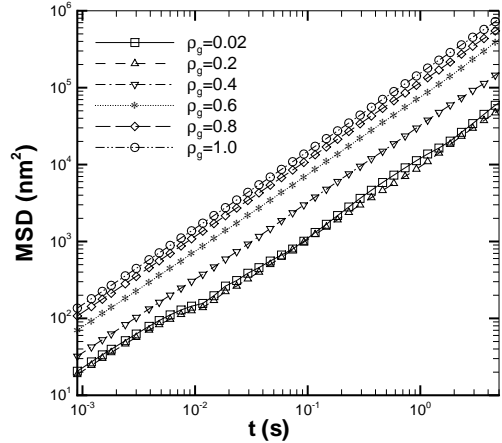}
\caption{Single Kinesin mean squared
displacement versus time for various densities. To a very good
approximation the diffusion is normal for densities
larger than $\rho_g \sim 0.2$ i.e.; the MSD grows linearly in time. } \label{fig:bz2}
\end{figure}

We have evaluated the slopes by linear curve fitting. To a very
good approximation the diffusion is normal for densities larger
than $\rho_g \sim 0.2$ i.e., the MSD grows linearly in time. It
would be more appropriate to define a line density $\rho$. This
quantity is defined as the time average number of bound Kinesins
to the MT. Note that a bound Kinesin can occupy two or one site
of the MT depending on whether one or two heads are attached. We
have sketched the dependence of $\rho$ on $\rho_g$ in Figure (9).

\begin{figure}
\centering
\includegraphics[width=7.5cm]{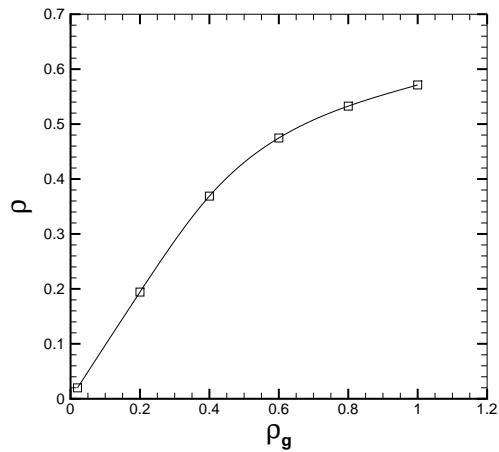}
\caption{Dependence of the line density $\rho$ on the global density $\rho_g$. $\rho$ tends to an asymptotic value near $0.7$. } \label{fig:bz2}
\end{figure}

We observe a linear increase of $\rho$ up to $\rho_g=0.4$.
Afterwards the rate of increase is lowered. This is due strong
repulsion between bound Kinesins. We stress that in order to have
values of $\rho$ larger than $0.6$ we have to significantly
increase $\rho_g$. This enormously rises the computational costs.
We extrapolated the behaviour of the curve for larger $\rho$. It
turns out that $\rho$ tends to an asymptotic value near $0.7$.
More specifically, let us denote the number of those bound
Kinesins attached via one head to MT by $N_1$ and those attached
via two heads by $N_2$ so we have $N=N_0+N_1+N_2$ ($N$ is the
total number of Kinesins). Furthermore, we show the number
unbound Kinesins by $N_0$ and the MT binding sites by $N_{bs}$.

\be \rho_g=\frac{N_0+N_1+N_2}{N_{bs}} ; ~~
\rho=\frac{N_1+N_2}{N_{bs}} \ee

In the extreme limit $\rho_g \rightarrow \infty $ we have (by
extrapolation) $\rho = 0.7$, on the other hand we have $2N_2
+N_1=N_{bs}$ which means all the sites are occupied by Kinesin
heads. We therefore find that sixty percent of the sites are
occupied by double-head attached Kinesins and forty percent by
single-head attached Kinesins. In the following graphs, we
exhibit the dependence of some cooperative quantities on the line
density $rho$. In Figure (10) we exhibit the dependence of a
single Kinesin diffusion coefficient $D$ versus $\rho$. It grows
slowly for small densities but shows a rapid increment for $\rho
> 0.3$. The reason is due to increase of unbound Kinesins when the
density is enhanced. Obviously, an unbound Kinesin performs a
diffusive motion therefore increasing the number of unbound
Kinesins gives rise to enhancement of the overall diffusion which
in turn leads to a larger $D$. Note that $D$ has been obtained by
averaging over the trajectories of all the Kinesins.

\begin{figure}
\centering
\includegraphics[width=7.5cm]{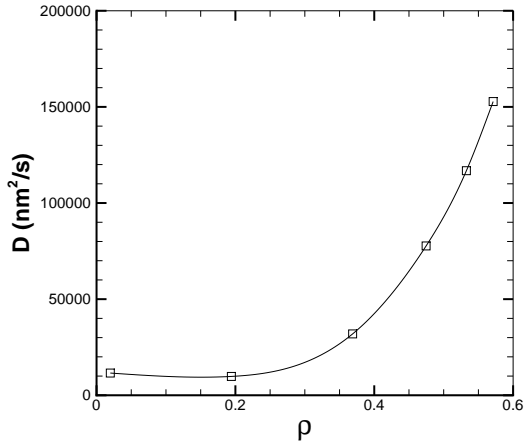}
\caption{Diffusion coefficient $D$ of a
single Kinesin versus density. The diffusion constant $D$ turns
out to be an increasing function of $\rho$. It grows slowly for
small densities but shows a rapid increment for $\rho > 0.3$. The
reason is due to increase of unbound Kinesins when
the density is enhanced. } \label{fig:bz2}
\end{figure}

We next turn into the issue of directed motion of Kinesins.
First, we have computed the dependence of the averaged velocity
of the CM versus $\rho$. Figure (11) sketches this
behaviour.

\begin{figure}
\centering
\includegraphics[width=7.5cm]{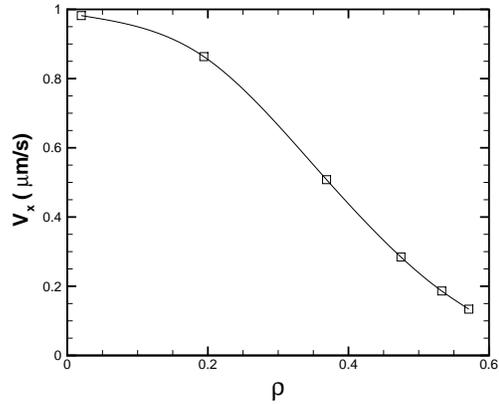}
\caption{Averaged CM velocity versus
density $\rho$. $\langle V_x \rangle$ is a decreasing function of
$\rho$. For small densities it decreases smoothly. When $\rho$ goes beyond $\rho=0.2$,
$\langle V_x \rangle$ shows a rapid decrease. } \label{fig:bz2}
\end{figure}

To obtain the above graph, we plotted $X_{cm}$ versus time and
fitted a linear curve to it. The slope of the fitted line
corresponds to the averaged velocity. As expected, $\langle V_x
\rangle$ is a decreasing function of $\rho$. For small densities
it decreases smoothly. However, when the density goes beyond
$\rho=0.2$, $\langle V_x \rangle$ shows a rapid decrease. In
figure (12) we have exhibited the mean directed passage length
(processive run length), the mean processive time and the
velocity of this directed motion on the MT as a function of
$\rho$.

\begin{figure}
\centering
\includegraphics[width=7.5cm]{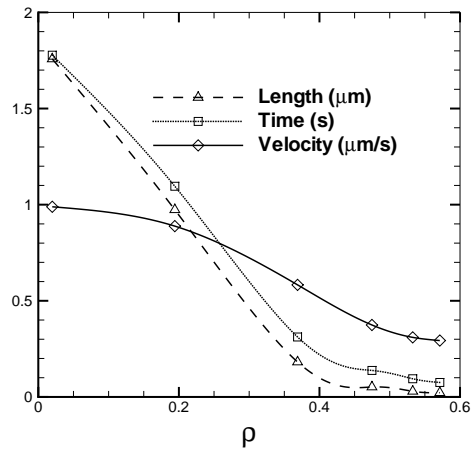}
\caption{Averaged directed processive
run length, time and velocity versus density. Average processive
run length exhibits a rapidly decreasing behaviour up to
$\rho=0.4$. After this value it shows a weak dependence on
$\rho$. The analogous behaviour is observed for the mean temporal
processivity. } \label{fig:bz2}
\end{figure}

The average processive run length $\overline{L}_{prs}$ defined as
the average distance a Kinesin can move proccesively on the MT
before detachment exhibits a rapidly decreasing behaviour up to
$\rho=0.4$. After this value it shows a weak dependence on
$\rho$. We speculate that at this density large traffic jams are
formed which do not allow Kinesins to move directedly. The
analogous behaviour is observed for the mean temporal
processivity $\overline{T}_{prs}$. Dividing $\overline{L}_{prs}$
by $\overline{T}_{prs}$ gives us the average velocity
$\overline{V}_{prs}$ of the processive motion. The average
velocity has a more smooth decreasing behaviour. Having obtained
the velocity of the directed motion, we are able to find the
dependence of the current of Kinesins along the microtubule. The
current $J$ is defined by the fluid mechanics relation $J=\rho_g
\overline{V}_{prs}$. Figure (13) sketches the dependence of $J$
versus $\rho$.

\begin{figure}
\centering
\includegraphics[width=7.5cm]{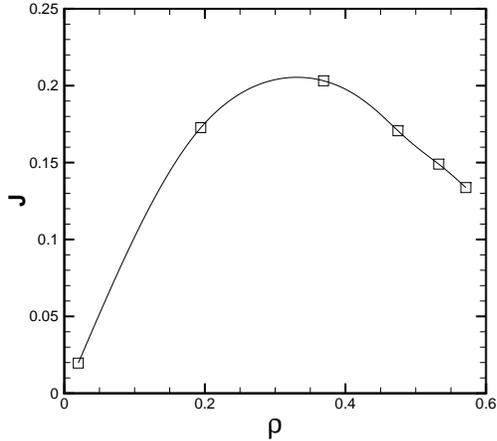}
\caption{Average Kinesins current versus density. It shows a maximum at $\rho \sim 0.34$. } \label{fig:bz2}
\end{figure}

$J$ exhibits a maximum at $\rho \sim 0.34$. The interesting point
is the existence of an asymmetry in the $J-\rho$ diagram.
Normally in lattice driven gas models such as {\it asymmetric
simple exclusion process} (ASEP) $J$ appears as a symmetric
function of $\rho$ \cite{asep}. The current $J$ is related to the
rate of cargo transport by Kinesin motor proteins. In order to
find a deeper understanding, we have evaluated the steady state
percentage of the bound and unbound Kinesins as a function of
$\rho$. Bound Kinesins are divided into two groups: one-head
attached and two-head attached to the MT. Figure (14) exhibits
this behaviour. Percentage of unbound Kinesins increases with density. In the
same manner, percentage of one-head attached Kinesins increases
with $\rho$ but it seems to become saturated at large densities.
Interestingly, percentage of two-head attached Kinesins shows a
decreasing behaviour. The rate of decrease is small at low
densities but becomes larger for higher densities.

\begin{figure}
\centering
\includegraphics[width=7.5cm]{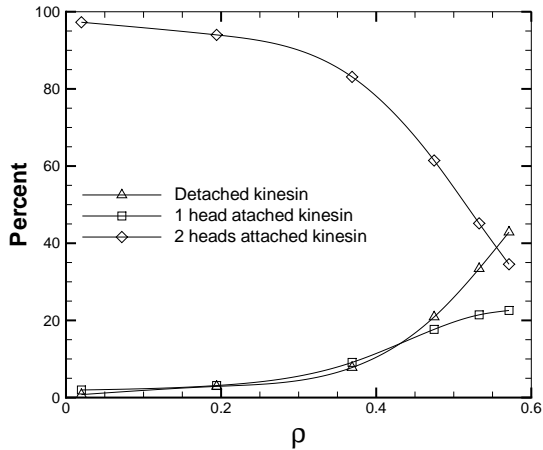}
\caption{Percentage of averaged bound and unbound Kinesins on the MT vs density. } \label{fig:bz2}
\end{figure}

\section{Summary and Concluding Remarks}

We have simulated the collective motion of Kinesins on a
microtubular track by developing a two dimensional Langevin-type
model. The model is capable of reproducing the hand-over-hand
mechanism of the directed motion along the microtubule. Various
quantities such as diffusion constant, average velocity and the
number current of Kinesins have been obtained by performing
extensive molecular dynamics simulations. Dependence of the above
quantities on the the overall density of Kinesins have been
obtained. It is shown that beyond a certain density, the average
velocity and the current undergo notable decrease which is due to
formation of traffic jams in the system. This observation is in
accord to the ubiquitous flow-density characteristics.

\end{document}